\documentclass[default,iicol]{sn-jnl}

\usepackage[version=4]{mhchem}
\usepackage{amssymb}
\usepackage[utf8]{inputenc}
\usepackage{siunitx}  

\newcommand{\abspsi}{\lvert \psi_6 \rvert}
\newcommand{\aveabspsi}{\left< \lvert \psi_6 \rvert \right>}
\newcommand{\aver}{\left< r_{sk} \right>}

\usepackage{xr}
\makeatletter
\newcommand*{\addFileDependency}[1]{
  \typeout{(#1)}
  \@addtofilelist{#1}
  \IfFileExists{#1}{}{\typeout{No file #1.}}
}
\makeatother

\newcommand*{\myexternaldocument}[1]{%
    \externaldocument{#1}%
    \addFileDependency{#1.tex}%
    \addFileDependency{#1.aux}%
}

\myexternaldocument{supplementary}

\jyear{2022}%

\begin{document}
\title{Thermal Hysteresis Behavior of Skyrmion Lattices in the van der Waals Ferromagnet \ce{Fe3GeTe2}}

\author[1,2]{\fnm{Arthur R. C.} \sur{McCray}}
\equalcont{These authors contributed equally to this work.}

\author[1]{\fnm{Yue} \sur{Li}}
\equalcont{These authors contributed equally to this work.}

\author[3]{\fnm{Rabindra} \sur{Basnet}}

\author[4]{\fnm{Krishna} \sur{Pandey}}

\author[3,4]{\fnm{Jin} \sur{Hu}}

\author[1]{\fnm{Daniel} \sur{Phelan}}

\author[5]{\fnm{Xuedan} \sur{Ma}}

\author[1,6]{\fnm{Amanda K.} \sur{Petford-Long}}

\author*[1]{\fnm{Charudatta} \sur{Phatak}}\email{cd@anl.gov}

\affil[1]{\orgdiv{Materials Science Division}, \orgname{Argonne National Laboratory}, \orgaddress{\city{Lemont}, \postcode{60439}, \state{IL}, \country{USA}}}

\affil[2]{\orgdiv{Applied Physics Program}, \orgname{Northwestern University}, \orgaddress{\city{Evanston}, \postcode{60208}, \state{IL}, \country{USA}}}

\affil[3]{\orgdiv{Department of Physics}, \orgname{University of Arkansas}, \orgaddress{ \city{Fayetteville}, \postcode{72701}, \state{AR}, \country{USA}}}

\affil[4]{\orgdiv{Materials Science and Engineering Program}, \orgname{University of Arkansas}, \orgaddress{ \city{Fayetteville}, \postcode{72701}, \state{AR}, \country{USA}}}

\affil[5]{\orgdiv{Center for Nanoscale Materials}, \orgname{Argonne National Laboratory}, \orgaddress{ \city{ Lemont}, \postcode{60439}, \state{IL}, \country{USA}}}

\affil[6]{\orgdiv{Department of Materials Science and Engineering}, \orgname{Northwestern University}, \orgaddress{\city{Evanston}, \postcode{60208}, \state{IL}, \country{USA}}}


\abstract{

Understanding the physics of phase transitions in two-dimensional (2D) systems underpins the research in diverse fields including statistical mechanics, quantum systems, nanomagnetism, and soft condensed matter. However, many fundamental aspects of 2D phase transitions are still not well understood, including the effects of interparticle potential, polydispersity, and particle shape. Magnetic skyrmions, which are non-trivial chiral spin structures, can be considered as quasi-particles that form two-dimensional lattices. Here we show, by real-space imaging using \textit{in situ} cryo-Lorentz transmission electron microscopy coupled with machine learning, the ordering behavior of N\'eel skyrmion lattices in van der Waals \ce{Fe3GeTe2}. We demonstrate a distinct change in the skyrmion size distribution during field-cooling, which leads to a loss of lattice order and an evolution of the skyrmion liquid phase. Remarkably, the lattice order is restored during field heating and demonstrates a thermal hysteresis. Our quantitative analysis explains this behavior based on the energy landscape of skyrmions and demonstrates the potential to control the lattice order in 2D phase transitions.

}



\maketitle

The organization and behavior of quasiparticle lattices is of fundamental importance when describing the complex properties of many condensed matter systems. For example, flux vortices in type II superconductors demonstrate both long- and short-range order as they undergo a solid-like to liquid-like phase transition at the superconducting transition temperature \cite{Zeldov1995, Safar1992}. Another magnetic quasiparticle of recent interest is the magnetic skyrmion, a topologically non-trivial spin structure that can exist both as individuals or in dense lattices \cite{Fert2017, Everschor-Sitte2018, Finocchio2016}. Skyrmions are most commonly stabilized by the Dzyaloshinskii-Moriya interaction (DMI)\cite{Dzyaloshinsky1958, Moriya1960}, with inversion-symmetry breaking intrinsic to a crystal structure giving rise to Bloch skyrmions, or inversion-symmetry breaking by interfaces creating N\'eel skyrmions.

The collective behavior of skyrmions is of great interest as a method for probing fundamental skyrmion properties and as a platform for studying two-dimensional (2D) phase transitions \cite{Huang2020}. The vast majority of work so far on skyrmion lattices has focused on Bloch skyrmions, which form well-ordered triangular lattices and display well-known dislocations and grain boundaries \cite{Huang2020, Mat2016, Pollath2017, Rajeswari2015}. N\'eel skyrmion lattices, by contrast, are much less studied and far more chaotic \cite{Zazvorka2020}, likely due to a weaker, and purely repulsive, interaction between the skyrmions \cite{Brearton2020, Capic2020}.

Here we study N\'eel skyrmion lattices in \ce{Fe3GeTe2} (FGT), a van der Waals (vdW) material that is ferromagnetic down to monolayer thickness and displays signatures that suggest enhanced electronic correlations \cite{Park2019, Zhu2016a, Fei2018, Leon-Brito2016}. FGT can be easily exfoliated, making it an appealing candidate for interfacing topologically-protected magnetic spin structures with superconductors \cite{Hals2016, Dahir2019, Menezes2019, Mascot2021, Rex2019, Chen2015}, topological insulators \cite{Chen2019a, Zhang2018a}, or magnonic materials \cite{Iwasaki2014, Psaroudaki2018}. The skyrmions observed in FGT are of particular interest as, unlike in conventional metallic multilayers, the skyrmion size is strongly dependent on temperature, thus forcing the lattice to adapt and realign as the temperature is changed \cite{Wu2020, Raju2019, Tomasello2018, Kwon2020}. By examining the order of N\'eel skyrmion lattices in response to varying temperature and magnetic field, we gain insight into how skyrmions interact with each other, are created and destroyed, and how the lattice itself evolves as a collection of dynamic skyrmions.

\section*{Magnetic Skyrmions in \ce{Fe3GeTe2}}\label{Sec.Res}
We performed \textit{in situ} cryo-Lorentz transmission electron microscopy (LTEM) experiments on skyrmion lattices in a 150 nm thick exfoliated FGT flake. We observe the skyrmion lattice response to field-cooling (FC) and field-heating (FH) processes under a range of applied field strengths, and quantified the skyrmion size and lattice order using AI-assisted analysis of the LTEM images.

\begin{figure*}[htb]%
\centering
\includegraphics[width=0.9\textwidth]{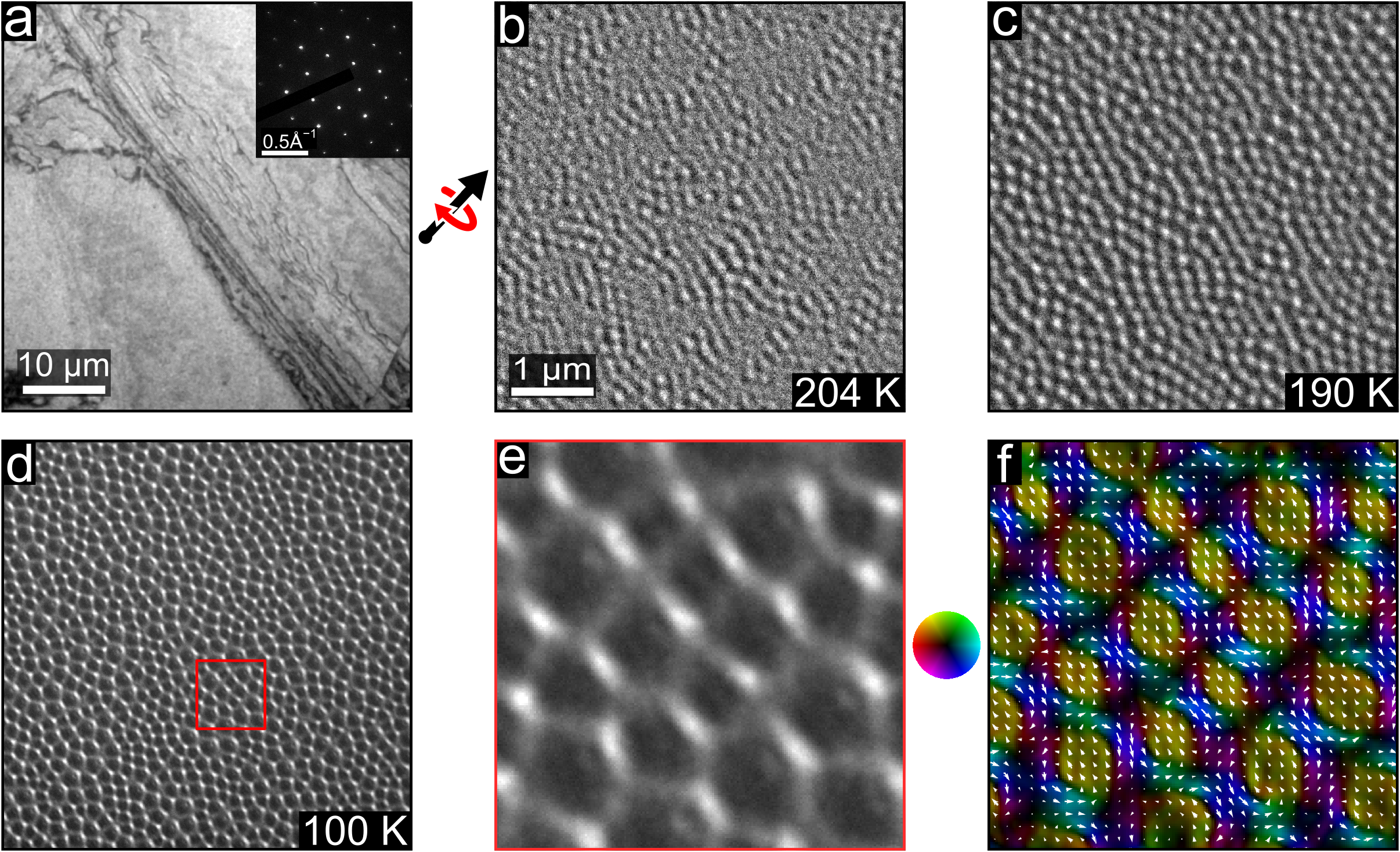}
\caption{\textbf{Magnetic skyrmions in \ce{Fe3GeTe2}.} \textbf{a} In-focus image of FGT flake. Inset: the [001] zone axis electron diffraction pattern. \textbf{b-d} Out-of-focus LTEM images at a defocus of $-7$ mm and sample tilt of $-23^\circ$, showing a skyrmion lattice during field cooling in a perpendicular 500 G applied field at \textbf{b} 204 K, \textbf{c} 190 K, and \textbf{d} 100 K. Images are at the same magnification and of the same region of the skyrmion lattice. \textbf{e} Magnified image of region highlighted by red box in \textbf{d}. \textbf{f} Map of the in-plane magnetic induction reconstructed from \textbf{e}; the yellow regions correspond to the positions of the skyrmion cores. Color wheel indicates magnetic induction direction and the black arrow indicates the tilt axis.}\label{Fig.Intro}
\end{figure*}

Figure \ref{Fig.Intro}a shows a low magnification image of an FGT flake recorded at room temperature with the inset showing the electron diffraction pattern for the [001] zone axis (i.e. the surface normal direction). Figure \ref{Fig.Intro}b-d show a sequence of images taken while field-cooling in a 500 G perpendicular applied field from above the observed Curie temperature (216 K) to a minimum temperature of 100 K. As the temperature was decreased, the average skyrmion size increased. Figure \ref{Fig.Intro}e shows the magnified image of a small region in Fig.~\ref{Fig.Intro}d highlighted by the red box. The corresponding  transport of intensity equation (TIE) reconstruction of the projected magnetic induction is shown in Fig.~\ref{Fig.Intro}f. Skyrmion lattices were formed when field-cooled with an applied field strength between 300 G and 750 G. Maze domains coexisted with skyrmions outside of this range (Supplementary Section S1).

\subsection*{Quantifying skyrmion lattice order with machine learning}
The location of each skyrmion in each image must be accurately determined in order to quantitatively analyze any changes to the skyrmion size and lattice order. The alternating bright/dark contrast of N\'eel skyrmions makes this difficult, especially when considering the decreased resolution caused by high defocus and the large field of view required to obtain statistically-significant results. We have therefore developed an automated method for determining skyrmion positions using a convolutional neural-network (CNN), which enabled us to identify more than 600,000 skyrmions across 380 images, giving robust statistical impact to our findings. As the skyrmions exist in a densely packed lattice, the center locations can be used to construct a Voronoi diagram and determine the radius of each skyrmion, $r_{sk}$. Details regarding the CNN and how $r_{sk}$ is calculated are discussed in the Methods section.

\begin{figure*}[htb]%
\centering
\includegraphics[width=\textwidth]{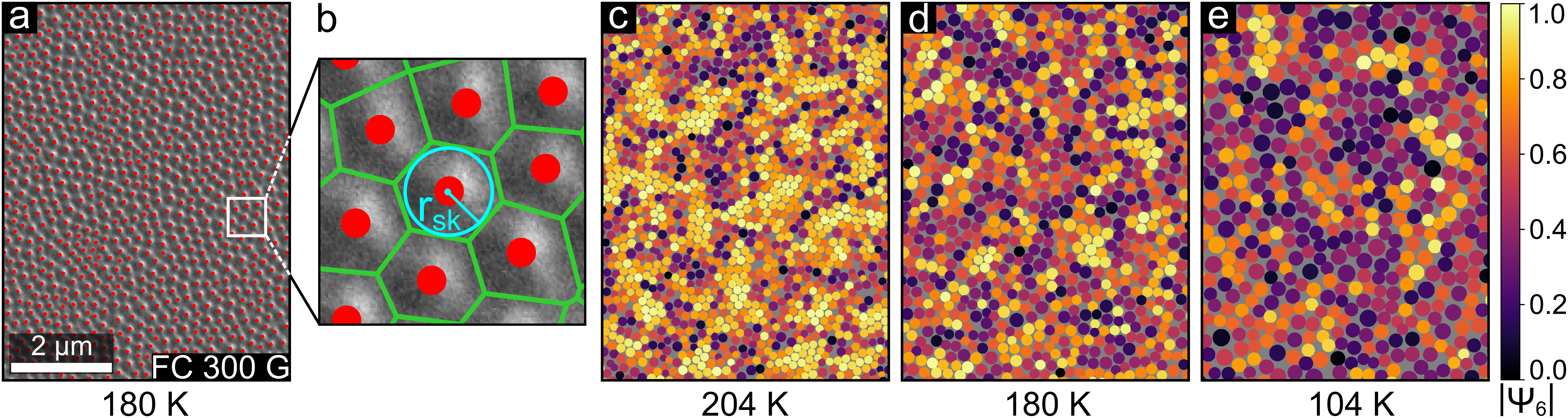}
\caption{\textbf{Skyrmion positions and skyrmion lattice order.} \textbf{a} LTEM image of a skyrmion lattice, with centers (red dots) identified using a CNN. \textbf{b} Enlarged region from \textbf{a}, showing the Voronoi diagram constructed from skyrmion centers as well as the skyrmion radius $r_{sk}$, determined as the radius of the largest inscribed circle within each Voronoi cell. \textbf{c-e} Colormaps showing circles of radius $r_{sk}$ at the position of each skyrmion center; color corresponds to the local 6-fold lattice symmetry as determined by $\abspsi$. Color scale indicated to right of images. Scale bar and field of view are identical to \textbf{a}. }\label{Fig.ML}
\end{figure*}

Figure \ref{Fig.ML} shows data from a FC sequence in a 300 G perpendicular applied field. In Fig.~\ref{Fig.ML}a, the LTEM image is overlaid with the skyrmion center locations identified by the CNN. The Voronoi diagram also allow us to define neighboring skyrmions as those that share a Voronoi cell edge, and with this we can calculate the local orientational order parameter \cite{Steinhardt1983, Mickel2013, kapfer2015} for each skyrmion,
\begin{equation}
    \psi_6 = \frac{1}{N} \sum_{j=1}^{N} e^{i 6 \theta_j},
\end{equation}
where $N$ is the number of neighbors, $\theta_j$ is the angle between the centers of the central skyrmion and its $j$th neighbor, and the summation is over all neighbors. The magnitude of $\psi_6$ is a measure of the local lattice symmetry around each skyrmion, with $\abspsi=1$ for a perfectly ordered lattice (see Supplementary Section S4).

Figure \ref{Fig.ML}c-e show qualitatively how skyrmion lattice order evolves when FC in an applied field of 300 G by visually mapping $\abspsi$ at three temperatures. A circle of radius $r_{sk}$ is plotted at each skyrmion location with the color corresponding to the local value of $\abspsi$. Stripes and clusters of local lattice order exist close to the Curie temperature (Fig.~\ref{Fig.ML}c), but lattice order decreases with temperature, and by 100 K there is very little sixfold coordination of the skyrmions within the lattice.

\section*{Hysteresis in the skyrmion lattice order}
\begin{figure*}[tb]%
\centering
\includegraphics[width=1.0\textwidth]{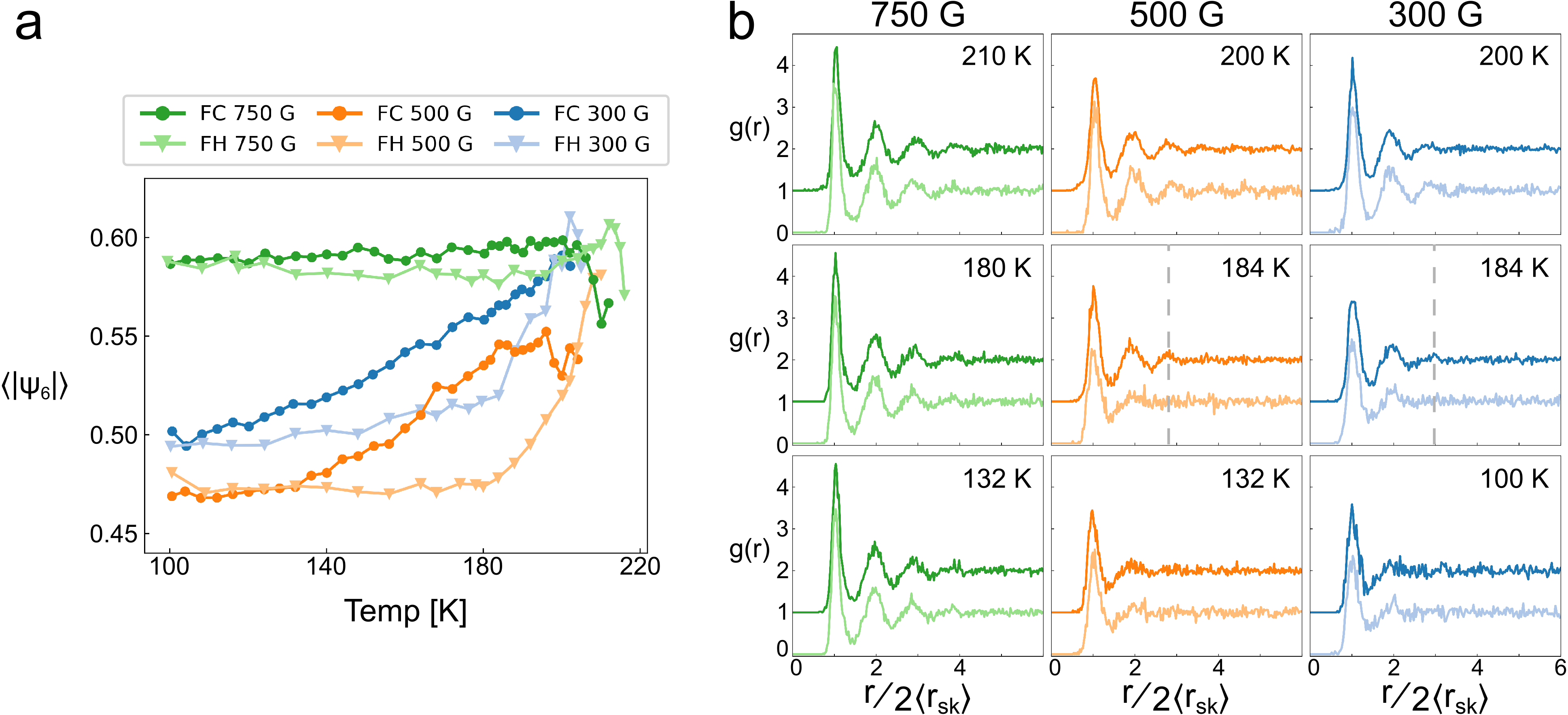}
\caption{\textbf{Hysteresis of skyrmion lattice order.} \textbf{a} Average value of $\abspsi$ across all skyrmions in a lattice, plotted as a function of temperature for FC and FH at a rate of $5$ K/min. \textbf{b} Radial distribution function, $g(r)$, plotted at three temperature points for each field cycle during FH (lighter curves) and FC (darker curves). $g(r)$ is plotted for the FH data and $g(r)+1$ is plotted for the FC data so that the curves do not overlap. All RDFs are plotted as a function of $r$ normalized by the average skyrmion diameter. }\label{Fig.Hyst}
\end{figure*}

We can average the values of $\abspsi$ across all skyrmions in an image to get $\aveabspsi$, a quasi-global measure of short-range lattice order \cite{Zazvorka2020, SampedroRuiz2019}. Figure \ref{Fig.Hyst}a shows $\aveabspsi$ as a function of temperature over three field-cooling and field-heating experiments carried out at three different field values. For the 750 G applied field, $\aveabspsi$ remains almost constant for both the FC and FH runs, indicating that the skyrmion lattice structure does not change significantly. However for the runs performed with fields of 300 G and 500 G, $\aveabspsi$ decreases with temperature during field cooling, indicating a loss of order that is restored upon field heating. For all three field values we observe a hysteresis that manifests as the decrease in lattice order while cooling not being restored until a higher temperature is reached during heating. This effect is minimal when the applied field is strongest, but for the 300 G and 500 G curves we observe a clear separation between the cooling and heating paths. The 500 G curve has the largest separation, with convergence between the FC and FH paths at 132 K and 200 K. 

Analyzing $\aveabspsi$ provides a fundamentally short-range metric to quantifying lattice order as it only accounts for each skyrmion's nearest neighbors. We therefore also look at the radial distribution function (RDF), $g(r) = \frac{1}{N} \sum_{\left< i,j \right>} \delta \left( \mathbf{r} - \mathbf{r}_{i,j} \right),$ of the skyrmion center locations, as shown in Fig.~\ref{Fig.Hyst}b for the same field values and at three different temperatures during FC and FH. For the data recorded at 750 G, we see relatively long-range order, indicated by the presence of 3-4 peaks in the RDF at all three temperatures when FC and FH. For the RDFs calculated at 500 G and 300 G there is a noticeable loss of order at lower temperatures shown by the single nearest-neighbor peak when both FC and FH. None of these RDFs show evidence of long-range order, and in even the most ordered lattices we only see short-range positional order limited at 5 peaks in the RDF (see Supplementary Section S5). 

The RDFs reinforce our findings of lattice order hysteresis as a function of temperature. The RDFs for an applied field of 500 G appear nearly identical when FC and FH at 200 K and 132 K, which coincides with the temperatures at which the FC and FH plots of $\aveabspsi$ converge. However, for the RDFs calculated at 184 K with a 500 G field, there is a clear additional peak observed during FC that is missing when the skyrmion lattice is field-heated, as highlighted by the dashed line. The loss of lattice order is therefore not confined to local six-fold symmetry, but also manifests as changes to the skyrmion locations at longer range. These data also show that the order regained at high temperatures includes both longer range translational effects (seen from the RDFs) and also local orientational order (quantified by $\aveabspsi$). The same hysteresis of the RDFs is visible in the data recorded at a field of 300 G, but to a lesser extent, which correlates with the less square hysteresis seen for $\aveabspsi$. In order to understand the origin of the decreasing lattice order and thermal hysteresis, we analyzed the variation of skyrmion size during the heating and cooling processes.

\section*{Skyrmion size variation}
\begin{figure*}[tb]%
\centering
\includegraphics[width=1.0\textwidth]{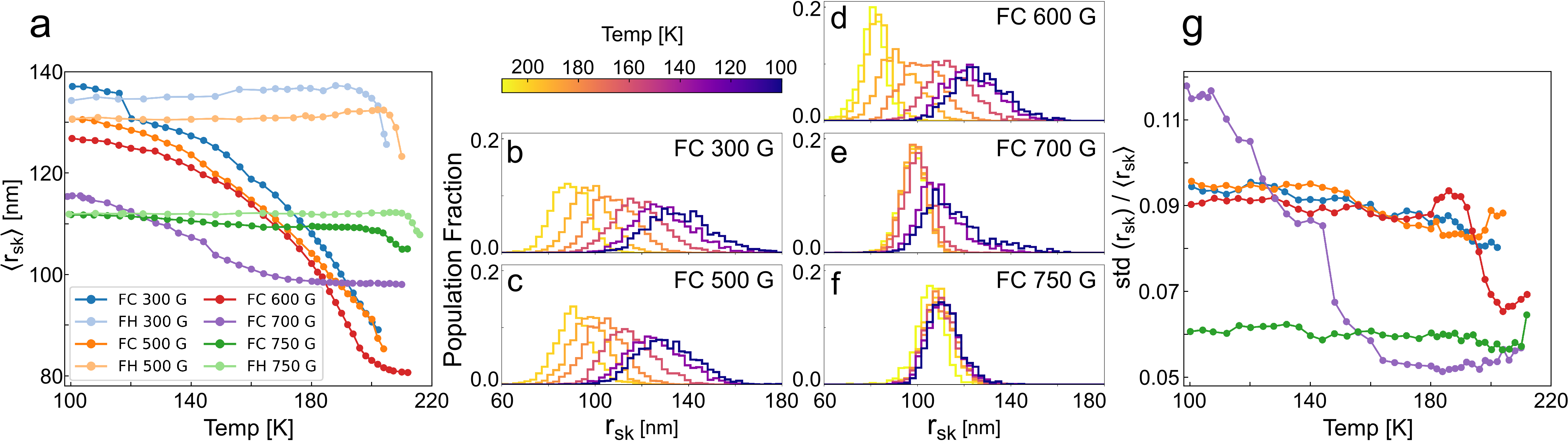}
\caption{\textbf{Change in skyrmion size with temperature.} \textbf{a} Average skyrmion size plotted as a function of temperature during FC for five different field values, and during FH for three different field values. \textbf{b-f} Histograms showing skyrmion size distribution as a function of temperature during FC for different applied field strengths. The histogram color denotes the temperature of the measurement according to the color scale. \textbf{g} Standard deviation of the skyrmion size distribution divided by the average skyrmion size as a function of temperature during FC for five field values; the plot legend is the same as in \textbf{a}.}\label{Fig.Sizes}
\end{figure*}

Figure \ref{Fig.Sizes}a shows the evolution of average skyrmion size $\aver$ as a function of temperature for FC and FH runs at five different field strengths. We first focus on the measurements during field cooling: for the 750 G applied field (green), the skyrmion size remains quite constant after initial nucleation throughout the explored temperature range. For a field of 700 G (purple) we observe that the skyrmion size is roughly temperature-independent at high temperature but sharply increases below 150 K. Similar behavior has been observed with weaker applied magnetic fields, but the skyrmion size starts to increase at higher temperatures. 
In particular, if the field is reduced to 500 G and lower (500 G: orange, 300 G: blue in Fig.~\ref{Fig.Sizes}), the skyrmion size increases upon cooling immediately after the formation of the uniform skyrmion lattice. In contrast, the situation is completely different when field-heating the sample. Under applied magnetic fields of 300 G (light blue), 500 G (light orange), and 750 G (light green), when the sample is heated up from 100 K, the average skyrmion size remains temperature independent up to the formation temperature of a uniform skyrmion lattice.
We note that there is a discontinuity in the measured skyrmion size for the 300 G field-cooled data close to 120 K. This is because we moved to a different region of the sample to avoid the bend-contour contrast obscuring the skyrmion contrast. We performed two additional FC experiments at 300 G and 500 G, which confirm the trends that we observe without the discontinuity (shown in Supplementary Section S2).

The observed growth of the average skyrmion size upon cooling is not caused by the continual deformation and growth of each skyrmion. Instead, we observe disappearance events wherein two or more skyrmions are replaced by one or more larger skyrmions.
The skyrmion disappearance events observed when cooling occur beyond the temporal resolution of our camera and therefore
cannot be directly resolved. Computational and analytical methods have shown that these events may occur via skyrmions merging, or by topological charge being ejected at the surface of the sample and nearby skyrmions expanding \cite{Birch2021, Lobanov2016}.

Analysis of the skyrmion size distributions shown in Fig.~\ref{Fig.Sizes}b-f provides a further explanation for how the skyrmion sizes change as a function of temperature. Each plot shows histograms of the skyrmion size distribution for all skyrmions in an image at representative temperatures through the FC process, with temperature indicated by the histogram color. There is little change in the narrow skyrmion size distribution for the 750 G applied field. However, for the lower field values the peak of the distribution shifts to higher values as temperature decreases, together with a noticeable increase in the width of the distribution. 

The histograms show that the skyrmion size distribution becomes wider as skyrmions grow, but they do not make clear how the width of the size distribution changes relative to the increased average skyrmion size. Figure \ref{Fig.Sizes}g plots the normalized width of the skyrmion size distribution, i.e. the standard deviation of skyrmion sizes divided by the average skyrmion radius, as a function of temperature, using the same colors as in Fig.~\ref{Fig.Sizes}a. The normalized skyrmion size distributions largely exist in two regimes. The lower region of the plot, corresponding to a narrower size distribution, contains the 750 G skyrmion lattices, the 700 G lattices above 150 K, and the 600 G lattices above 190 K. This regime depicts skyrmion lattices where skyrmions are in an as-nucleated state and have sizes that are stable as a function of temperature, as shown in Fig.~\ref{Fig.Sizes}a. The remaining 300 G, 500 G, and 600 G (below 190 K) skyrmion lattices all exist in a second regime with wider skyrmion size distributions, and the average skyrmion size in these lattices increases with decreasing temperature.

Fig.~\ref{Fig.Sizes}a and g show that the skyrmion lattices nucleate near the Curie temperature with a relatively narrow size distribution, but this distribution widens when cooling as skyrmion growth events occur. Within the growth-event temperature regime, the normalized width of the skyrmion size distribution remains temperature-independent. Additionally, the strength of the applied field affects the average skyrmion size but does not affect the normalized width of the size distribution. The 700 G data (purple) below 150 K are an exception to this trend, as we observe a much larger size variation in these skyrmion lattices. This is due to skyrmion growth events occurring non-uniformly across the skyrmion lattice as discussed in Supplementary section S3. 

We can conclude from Figs.~\ref{Fig.Hyst} and \ref{Fig.Sizes} that the increase in skyrmion size and in the width of the size distribution are what lead to a loss of skyrmion lattice order. This is not surprising, as it is much easier to tile circles of uniform size than to tile circles with a large size distribution. Regardless of the particular mechanism by which skyrmion growth events occur, it is clear that the sudden appearance of large skyrmions plays a major role in destroying the lattice order. 

\section*{FGT energy landscape}
The change in skyrmion size with temperature, and the dependence on applied field strength, can be explained by considering the energy landscape of the skyrmions in a micromagnetics framework. The relative strengths of the exchange, anisotropy, DMI, dipolar, and Zeeman energies are what initially cause the skyrmions to form at the Curie temperature, and they also determine the most stable skyrmion size.

\begin{figure}[tb]%
\centering
\includegraphics[width=\columnwidth]{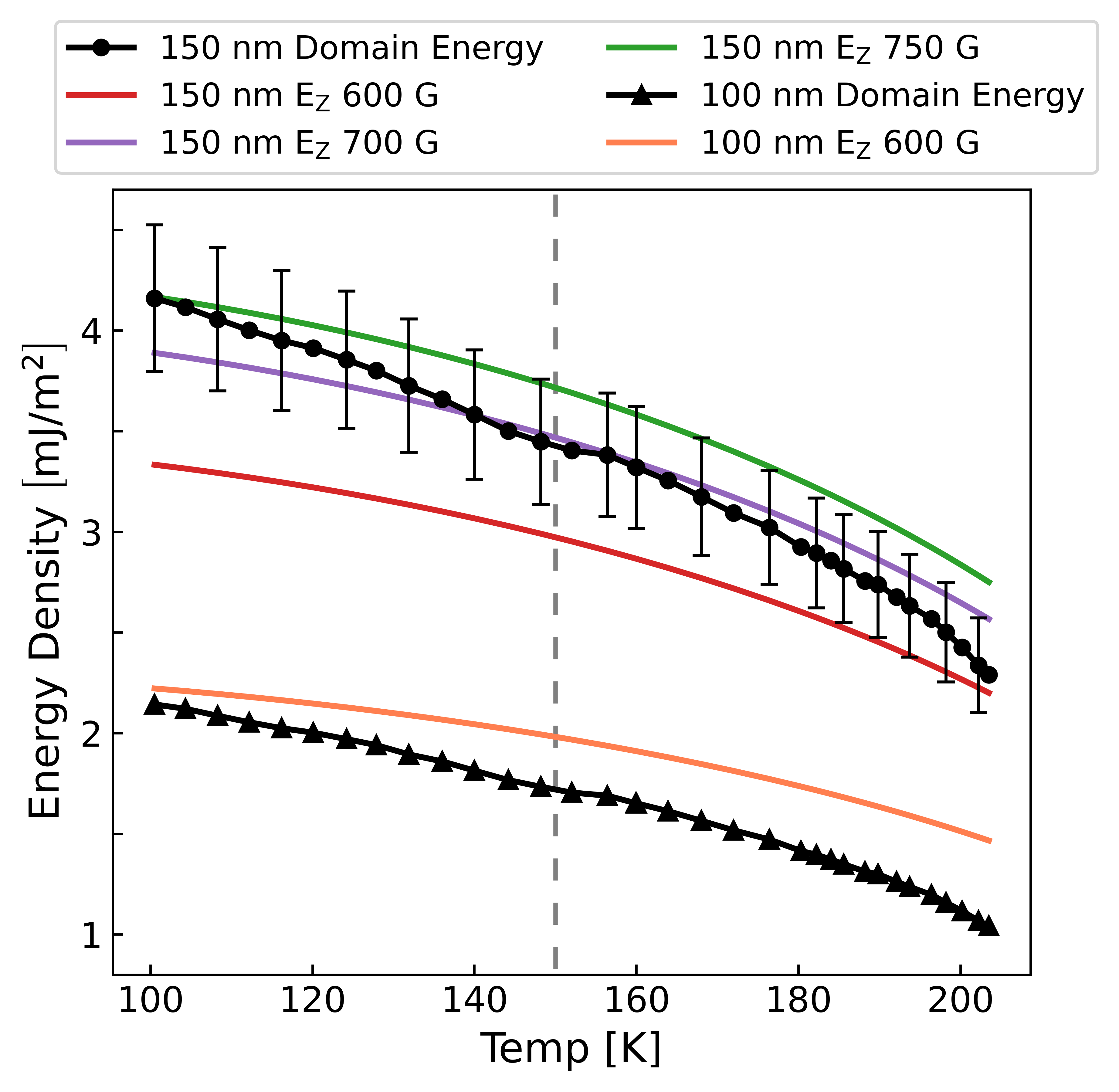}
\caption{\textbf{Skyrmion domain energy and Zeeman energy vs temperature.}  Skyrmion lattice domain energy density calculated from FGT magnetic parameters and observed domain size, plotted with Zeeman energy as a function of temperature. Error bars show one standard deviation of uncertainty and grey dashed line marks 150 K, where FC skyrmions at 700 G begin to increase in size.} \label{Fig.Energy}
\end{figure}

These energy terms are all dependent on the magnetic parameters of FGT, which are themselves highly temperature dependent. We performed several experimental measurements of magnetization, hysteresis loops, and domain width in FGT as a function of temperature. Through a combination of these experimental measurements and micromagnetic simulations, we parameterized the magnetic properties of FGT across our temperature range (see Supplementary Section S7). These parameterized properties were used, together with experimental measurements of skyrmion size, in a domain wall model \cite{Meier2017} to calculate the skyrmion domain energy density (details in Supplementary Section S8). This energy density is plotted as a function of temperature (connected black dots) in Fig.~\ref{Fig.Energy} along with the Zeeman energy arising from the external applied field (colored lines) for our 150 nm thick sample. The corresponding plot for a 100 nm thick sample shows that both the energies are lowered as expected. The Zeeman energy term increases as the temperature decreases, as it is proportional to the saturation magnetization, but the domain energy increases more sharply. We conclude that the dominant energy term can be either the skyrmion domain energy or the Zeeman energy, depending on the specific combination of temperature and field. As shown in Fig.~\ref{Fig.Energy}, our model shows that the crossover point occurs near 150 K when a 700 G field is applied, which agrees with the temperature at which the average skyrmion size starts to increase (Fig.~\ref{Fig.Sizes}a). Likewise, for a 750 G applied field the skyrmion size is stable well below the Curie temperate as the Zeeman energy is consistently greater than the domain energy. The energy terms and domain energy do not plateau until 50 K, which is below our experimental temperature range. 

The energy landscape also helps us discuss the return of lattice order that is seen when heating the samples back to the Curie temperature. The larger skyrmions that form at low temperatures are highly metastable, with their size limited by the confining influences of the surrounding skyrmions, and they do not shrink until nearly reaching the Curie temperature (see Supplementary Video 1). The return of lattice order cannot therefore be explained by the lattice returning to the original domain state that nucleates at the onset of field cooling. Rather, it appears that at low temperature, the skyrmions are frozen or held in place depending on where the skyrmion growth events occur. 

This is supported by observations in a 100 nm thick FGT sample, in which we did not observe pronounced changes in skyrmion size. As seen in Fig.~\ref{Fig.Energy}, the 100 nm thick sample shows reduced energy density (connected black triangles) relative to the Zeeman energy for a 600 Oe applied field (orange line). This indicates that a lower strength field, perhaps even as low as is necessary to stabilize skyrmions, would prevent the skyrmion size changing over our measured temperature range.

\section*{Skyrmion lattice phase and outlook}\label{sec.Discussion}
Studies of skyrmion lattices and 2D phase transitions often distinguish between systems existing in crystalline, hexatic, and liquid phases \cite{Nelson1979, Bernard2011, kapfer2015}. Previous studies have shown that Bloch skyrmion lattices can exist in all three phases \cite{Huang2020}. Although N\'eel skyrmion lattices have been predicted to have a hexatic phase \cite{Zazvorka2020}, we have not observed this. We found no indication of long-range positional or orientational order in the radial distribution functions, orientational correlation functions, structure factors, or any other metrics (see Supplementary Section S5). Therefore, while we have shown that the N\'eel skyrmion lattices do display controllable degrees of both positional and orientational order, all of the lattices that we have observed are confined to the skyrmion liquid phase.

One potential explanation for the different organization of Bloch and N\'eel skyrmion lattices comes from the origins of the DMI. Single crystal samples that show Bloch skyrmions have uniform energy terms across the entire sample, leading to uniformly sized skyrmions. In samples that exhibit N\'eel skyrmions, by contrast, the energy terms and the DMI strength may vary across the sample due to factors such as layer roughness and strain especially in exfoliated flakes. N\'eel skyrmion lattices therefore do not show the uniform size of Bloch skyrmions and, lacking any long-range interactions or sufficiently strong short-range interactions, this variation in skyrmion size may make it is impossible to achieve a hexatic or crystalline phase \cite{Russo2017}. Our work provides a platform for understanding the collective behavior of skyrmions by focusing on how interactions and fluctuations affect the 2D skyrmion lattice order. Future research using fast-imaging could provide insights into how and why skyrmions are destroyed and grow, as well as study how strain and interfacial effects in the vdW material can be used to control the skyrmion lattice.

\bibliographystyle{sn-standardnature}
\bibliography{References_final.bib}

\section*{Methods}\label{sec.Methods}
\subsection*{Sample preparation.}
A chemical vapor transport technique was used to grow millimeter-sized FGT single crystals. An evacuated quartz tube containing a stoichiometric mixture of Fe, Ge, and Te powders, along with with a transport agent \ce{I_2}, was heated in a two-zone tube furnace for 1 week with a temperature gradient of 700 to \SI{600}{\celsius}. The compositions of the crystals were determined by energy-dispersive x-ray spectroscopy.

A bulk FGT crystal was exfoliated mechanically using tape in an ambient atmosphere. Flakes were picked up using a polydimethylsiloxane (PDMS) stamp and transferred to the window of a silicon nitride TEM chip by heating it to approximately \SI{80}{\celsius}. The average thickness of the FGT flake was measured to be approximately 150 nm using an atomic force microscope.

\subsection*{LTEM measurements.}
Field cooling measurements were performed in a JEOL JEM-2100F TEM instrument in Lorentz mode using a Gatan liquid nitrogen holder. In order to image N\'eel domain walls and N\'eel skyrmions using LTEM, the image must be defocused and the sample tilted. The N\'eel nature of the domain walls and skyrmions was confirmed because magnetic contrast was seen at tilt angles of $+20^\circ$ and $-20^\circ$, but was not visible at $0^\circ$ \cite{Jiang2019}. Although FGT has been reported to host a variety of Bloch and N\'eel magnetic domain structures \cite{Park2019, Ding2020, Wu2020}, we have observed purely N\'eel domains and skyrmions.

A  perpendicular magnetic field of strength 0 to 1350 G was applied and controlled by the objective lens current in the microscope. When field-cooling, the sample was initially cooled to 240 K and allowed to stabilize before cooling proceeded to 100 K at a rate of -5 K/min. Images were taken at a nominal defocus of -7 mm, with alpha-tilt values of approximately $-20^\circ$, and were recorded on the same sample regions except where noted. Both the sample position and tilt value were adjusted slightly to avoid diffraction contrast, which changed with each cooling/heating cycle. Upon reaching 100 K the sample was allowed to rest for 10 minutes before field heating began. We did not observe any changes to the skyrmion lattice when holding at 100 K for up to 40 minutes. Field heating was also performed at a rate of 5 K/min, with imaging conditions and sample location identical to those used during field cooling.

During cooling, magnetic domains were first observed at 216 K, but the onset was non-uniform with skyrmions appearing in isolated clusters that later grew together to form an extended lattice. The non-uniform onset of ferromagnetic order is likely due to small compositional variations in the sample or due to strain that is induced when cooling the sample on a SiN membrane \cite{Zhu2021, Hu2020}. Upon cooling to 200 K, skyrmions could be identified across the $10\ \mathrm{\mu m}$ field of view, with a uniform skyrmion lattice forming at a relatively higher temperature for higher applied fields. Analysis of skyrmion lattice order and skyrmion sizes was performed only for images for which the skyrmion lattice was visible across the field of view. The skyrmion disappearance events observed when cooling occur beyond the temporal resolution of our camera and therefore the mechanism by which some skyrmions are destroyed and others grow cannot be directly resolved.

\subsection*{Machine learning for skyrmion finding.}
Image data was filtered to remove hot/dead pixels, then intensity was normalized, and image size was unidirectionally scaled to account for sample tilt before being input to the CNN. The CNN is implemented in PyTorch\cite{Pytorch2019}, and is a fully convolutional neural net based on a U-Net model, which has proven robust for segmenting images across applications ranging from biological imaging to atomic-resolution scanning TEM \cite{Ronneberger2015, Ziatdinov2017}.  A set of 6 $2048 \times 2048$ images were selected for which the skyrmions were identified using a combination of traditional image analysis techniques and checked by hand. These six images were then augmented and noise was added to generate a training data set consisting of 3000 $256\times256$ images which were used to train the CNN.

\subsection*{Quantifying skyrmion size from center locations.}
The current version of our neural net identifies only skyrmion centers and not the absolute size of each skyrmion. Because the skyrmions exist in a densely-packed lattice, we can determine the skyrmion size by creating a Voronoi tessellation based on the skyrmion centers. Each skyrmion then exists inside a Voronoi cell containing all points closer to its center than to any other skyrmion center. For each skyrmion cell we calculate the largest inscribed circle and define the radius of this circle, $r_{sk}$, to be the skyrmion size, as shown in Fig.~\ref{Fig.ML}b. Using the inscribed circle method is superior to simply averaging the nearest neighbor distances or using the Voronoi cell area, because it is not skewed by individual neighbors being displaced or by irregularly spaced skyrmions. This method has been verified against micromagnetic simulations of skyrmion lattices, and the skyrmion radii are very similar to those found directly from the magnetization (see Supplementary Section S4). 

\backmatter

\bmhead{Data availability} The raw data that support the findings of this study are available from the corresponding author upon reasonable request.

\bmhead{Code availability} The machine learning implementation for skyrmion finding as well as a trained CNN is available online at \url{https://github.com/Art-MC/SkX_NN}. 

\bmhead{Acknowledgments}
We thank O. Heinonen for helpful discussions. This work was supported by the U.S. Department of Energy, Office of Science, Office of Basic Energy Sciences, Materials Sciences and Engineering Division. Use of the Center for Nanoscale Materials, an Office of Science user facility, was supported by the U.S. Department of Energy, Office of Science, Office of Basic Energy Sciences, under Contract No. DE-AC02-06CH11357. Single crystal growth and magnetization measurements up to 9T is supported by the U.S. Department of Energy, Office of Science, Basic Energy Sciences program under Grant No. DE-SC0022006. This research used resources of the Argonne Leadership Computing Facility, which is a DOE Office of Science User Facility supported under Contract DE-AC02- 06CH11357.

The submitted manuscript has been created by UChicago Argonne, LLC, Operator of Argonne National Laboratory (“Argonne”). Argonne, a U.S. Department of Energy Office of Science laboratory, is operated under Contract No. DE-AC02-06CH11357. The U.S. Government retains for itself, and others acting on its behalf, a paid-up nonexclusive, irrevocable worldwide license in said article to reproduce, prepare derivative works, distribute copies to the public, and perform publicly and display publicly, by or on behalf of the Government. The Department of Energy will provide public access to these results of federally sponsored research in accordance with the DOE Public Access Plan. http://energy.gov/downloads/doe-public-access-plan

\bmhead{Author contributions}
RB and KP grew the single crystal samples and carried out magnetometry measurements of hysteresis loops of the crystals. DP performed magnetometry measurement for temperature-dependent curves of bulk crystals. YL prepared samples for LTEM imaging, analysed magnetometry data and performed micromagnetic simulations. ARCM and YL performed LTEM imaging. ARCM analyzed the data and wrote the manuscript. Both ARCM and YL contributed equally and have the right to list their name first in their CV. All authors discussed the results and commented on the manuscript. 

\bmhead{Competing interests}
The authors declare no competing interests.  

\bmhead{Supplementary information} Supplementary information is available for this paper as a separate  PDF.

\end{document}